# How Ground Deformation Influences Earthquake Occurrence During the Ongoing Unrest at Campi Flegrei (2005-Present)

Cataldo Godano[1*], Vincenzo Convertito[2], Anna Tramelli[2] and Giuseppe Petrillo[3]

[1]Università della Campania "L. Vanvitelli"
Dipartimento di Matematica e Fisica
Viale Lincoln 5 - 81100 Caserta - Italy

[2]Istituto Nazionale di Geofisica e Vulcanologia,
Sezione di Napoli - Osservatorio Vesuviano -
Via Diocleziano, 328, 80125 Napoli

[3]Earth Observatory of Singapore (EOS),
Nanyang Technological University (NTU)
50 Nanyang Ave, Block N2-01a-15, Singapore 639798

*Corresponding author: cataldo.godano@unicampania.it

March 26, 2025

**Abstract**

We investigate the relationship between the cumulative number of earthquakes and ground uplift at the Campi Flegrei caldera (South Italy) during the ongoing unrest (2005-present). While previous studies have explored this correlation, we propose a nonlinear epidemic model that captures new features of the caldera system. Our model describes earthquakes' occurrence as a cascading process driven by ground deformation. The nonlinearity reflects the reduced efficiency of the triggering mechanism, which contributes to the short duration of seismic swarms. This mechanism may represent a general framework for understanding the occurrence of volcanic earthquakes worldwide.

## 1 Introduction

The Campi Flegrei volcanic field has sometimes been described in journalistic contexts as a "super-volcano" due to its extension ($\simeq 450 km^2$) and to the two huge main eruptions that left the caldera topography: the Campanian Ignimbrite (39000 years ago) and the Tufo Giallo Napoletano (15000 years ago) [1, 2]. Following the Tufo Giallo Napoletano eruption, at least 70 intracaldera eruptions occurred [1]. The most recent of these marked the end of the eruptive sequence in 1538 AD, leading to the formation of Monte Nuovo.

Periods of subsidence and uplift of the soil have been observed since Roman times. After the Monte Nuovo uplift and eruption, Campi Flegrei experienced, in historical times, 4 unrest episodes: 1950-1952, 1970-1972, 1982-1984 and 2005-now day with a total uplift of 4.3 m [3–6].

As for the 1950-1952 unrest, due to the limited capacity of the surveillance system, few data are available. However, felt seismicity was negligible [3, 4]. The 1970-1972 unrest was characterized by moderate seismicity accompanied by a soil uplift in the central part of the caldera of about 1.5 m [3, 4].

The following unrest episode started on 2 November 1982 and ended on December 1984 accompanied by about 16000 earthquakes and a soil uplift of approximately 1.8 m [7]. The seismicity occurred mainly in the Solfatara - Pozzuoli area at a depth shallower than 3 km, whereas deeper events occurred off-shore [7, 8]. The extensional regime, in agreement with the inferred stress field, was enlightened by the focal mechanisms [9]. Different approaches and models have been suggested for explaining the very localized Gaussian-shaped soil uplift [10]. Since 1960 it has been very difficult to provide a mechanical model explaining the observed soil deformation [11] leading to the suggestion of many different mechanisms that can also mix together: a very shallow and highly over-pressured source, more than one source at different depths, lens-shaped magmatic body, the effect of the caldera boundary faults, (see Bonafede et al. [12] for a comprehensive review).

Differently from the 1982-1984 , the present unrest started in 2005 with slowly increasing soil uplift. However, the first significant earthquake swarm was observed in October 2006 and it was necessary to wait approximately 12 years to observe a relevant number ($\simeq$500) of seismic events in the area. This implies that the stress (magnitude and orientation) induced by the source of the soil uplift on the active faults system was not still sufficient to generate a large number of earthquakes [13]. The ground uplift started in 2005 is now-day still increasing and appears to be correlated with the seismic activity (see section 2 for a more detailed description). Moreover, many other observables (e.g. $CO/CO_2$, inter-event time between successive earthquakes) seem to be correlated one with each other suggesting a common source [6] for temporal variations of these observables.

It has been suggested that the whole sequence of unrest since 1950 could be interpreted as a single process of crustal extension [14–16] under the action of magmatic fluid pressurization. More precisely, the whole unrest sequence could be explained in terms of the mechanical behaviour of the upper crust transitioning from purely elastic, to quasi-elastic and, finally, to inelastic.

Here we investigate the correlation between the ground vertical deformation and the number of earthquakes and interpret this observation by means of a very simple model assuming a non linear dependence of the cumulative earthquake number on the deformation.

## 2 The soil uplift

The vertical soil displacement $\delta$ recorded at the GNNS station RITE managed by the INGV-Osservatorio Vesuviano, can be divided into four different periods (Fig. 1). The first one (2000-2005) is characterized by a $\delta$ decrease and precedes the unrest starting in 2005 when the soil uplift increase is marked by a deformation peak ending in 2007 (second period 2005-2008). The third (2008-2013) and fourth (2013-2024) periods are characterized by two successive exponential increases. The breaking point occurred at the beginning of 2013 when the slope of the curve changed. In September 2012 a seismic swarm of 188 events occurred in a day ($m_{max}$ = 1.7) during a period when the earthquake occurrence rate was less than 20 events per month. The swarm was preceded by an uplift velocity increment, followed by a decrease of the soil deformation and accompanied by an increment of the hydrothermal activity (new opening of fumarolic vents) [17]. In spite of the smaller slope of the curve in period 4, the velocity of the increase of $\delta$ is larger in this period (see inset of Fig. 1) because the coefficient of the exponential fitting curve is one order of magnitude larger than the one relative to period 3.

Bevilacqua et al. [18] suggested a parabolic behaviour of the soil uplift time dependence using part of period 2, period 3 and period 4. In fact, a different fit could be attempted by joining the third and fourth periods. In this case, $\delta$ increases with time as a power law (see Fig. 2). However, the $\chi^2$ for the power law is equal to 17.6, whereas $\chi^2$ = 3.6 for the double exponential. This makes the fit of Fig. 1 statistically preferable to the one of Fig. 2. Moreover, the changing point separating the two exponential regimes coincides with the occurrence of the 2012 earthquake swarm, which likely signs a change in the dynamic of the caldera [19, 20].

The existence of a correlation between the cumulative number of earthquakes $N$ and soil uplift $\delta$ was already enlightened by Kilburn [14], Kilburn, De Natale, and Carlino [15], and Kilburn et al. [16] who suggested exponential increases of $N$ with $\delta$ followed by a linear dependence. Conversely, Bevilacqua et al. [18] suggested a pure exponential fit for the dependence of the cumulate number of earthquakes $N$ on soil uplift $\delta$. However, the figure 3 reveals that an exponential fit does not represent a very good choice and that a linear regime, following the exponential one, is not observable.

The red line in figure 3 represents the fit of the exponential model:

$$N = (35.25 \pm 0.005)e^{(0.035 \pm 9 \cdot 10^{-5})\delta} \tag{1}$$

with $\chi^2 = 6.9 \cdot 10^{-3}$.

Notice that the first 30 earthquakes have been excluded from the analysis because they appear to be $\delta$ independent.

Here, we suggest a more appropriate expression:

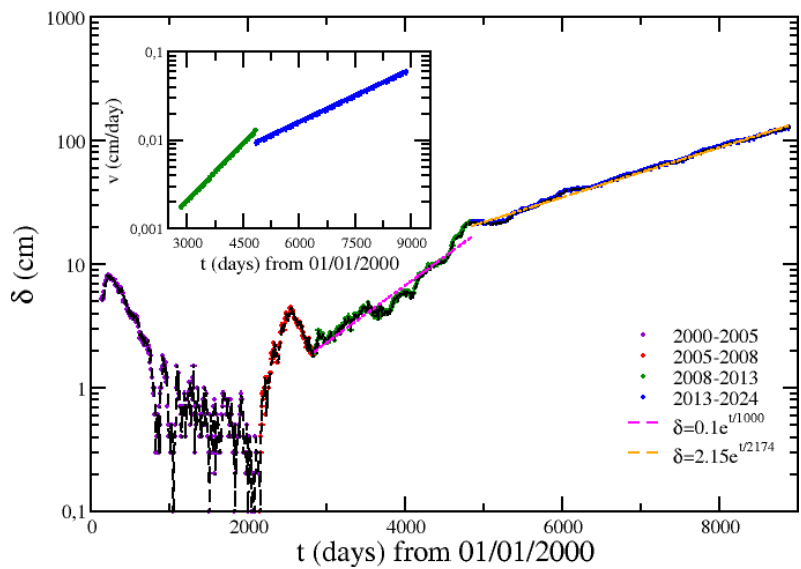

Figure 1: $\delta$ as a function of time on a semi-log scale. The dashed lines represent the two exponential fits obtained for periods 2008-2013 and 2013-2024. The inset shows the time derivative (velocity) of the obtained fits.

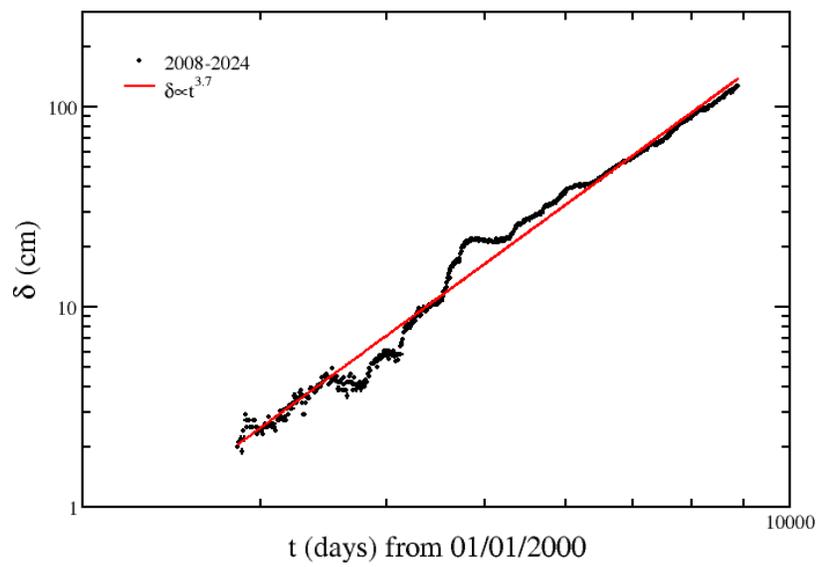

Figure 2: $\delta$ as a function of time on a log-log scale. The red line represents the power law fits obtained joining periods 2008-2013 and 2013-2024.

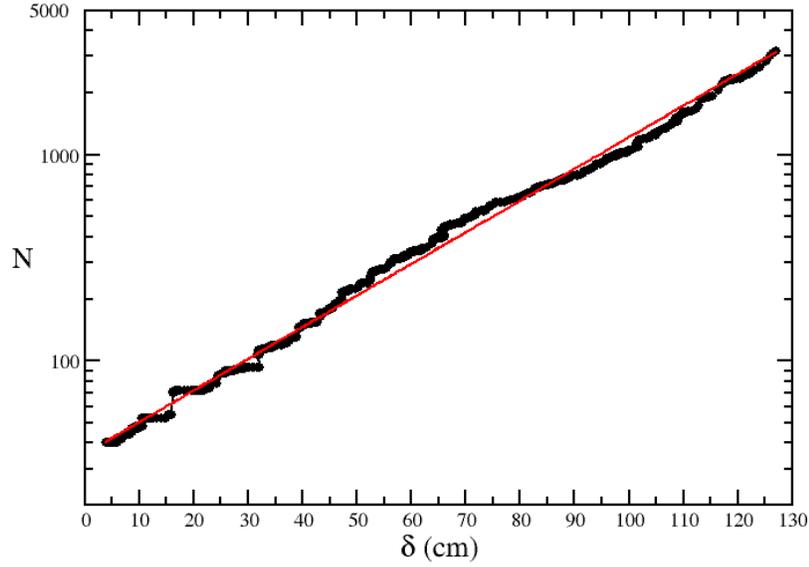

Figure 3: The cumulative number of earthquakes as a function of $\delta$ on a semi log scale. The red line represents an exponential fit.

$$\ln(\ln(N(\delta))) = k_1 \delta^{k_2} \tag{2}$$

The meaning of the double logarithm will be explained in the following section.

Fig. 4 shows $\ln(\ln(N(\delta))$ as a function of $\delta$ on a log-log scale.

The red line is the power law fits:

$$\ln(\ln(N)) = (0.76 \pm 0.003)\delta^{0.2 \pm 0.00067} \tag{3}$$

with $\chi^2 = 5.6 \cdot 10^{-5}$. An $F$-test reveals that the power-law fit must be preferred at a 95% significance level.

In what follows, we provide a simple model for explaining the fit obtained.

## 3 The model

The epidemic nature of earthquake occurrence has been recognized since the early 1970s [21–25], leading to the development of the Epidemic-Type Aftershock Sequence (ETAS) model. The ETAS model and its subsequent refine-ments [26–34] have been instrumental in describing the temporal and spatial clustering of earthquakes due to the triggering of aftershocks.

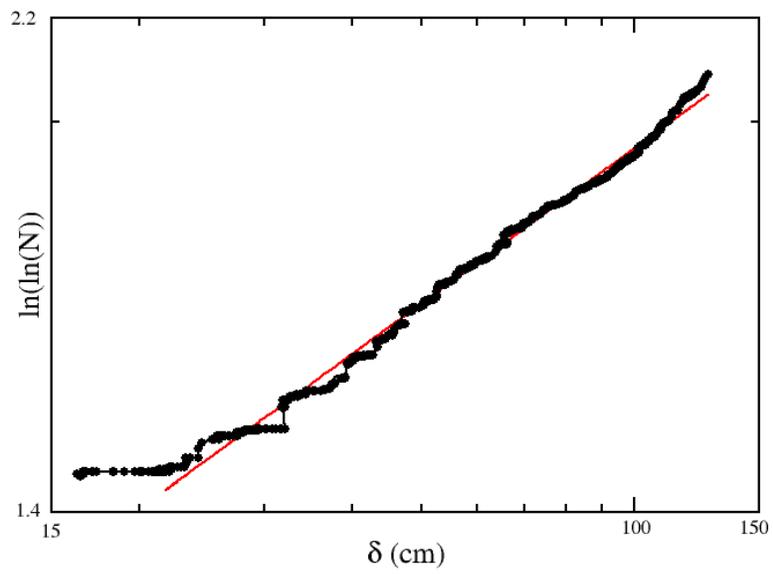

Figure 4: The double logarithm of the cumulative number of earthquakes as a function of δ on a log-log scale. The red line represents the power law fit of Eq. 2.

In this study, we adopt an epidemic-like equation to model the dependence of the earthquakes' occurrence rate on ground uplift $\delta$. The use of $\delta$ as the primary variable reflects its role as a proxy for fluid pressurization and stress/strain accumulation in volcanic settings. Unlike time-based models, which are more suitable for tectonic sequences, $\delta$ directly represents the physical processes that drive seismicity, making it a more appropriate choice for volcanic swarms. Moreover, it captures the cumulative deformation over time.

We propose the following differential equation to describe the rate of earthquake occurrence with respect to ground uplift:

$$\frac{dN}{d\delta} = \beta \delta^\alpha N \ln(N), \tag{4}$$

where $N(\delta)$ is the cumulative number of earthquakes at a given ground uplift $\delta$, $\beta$ is a proportionality constant, and $\alpha$ characterizes the non-linear dependence on $\delta$.

## 3.1 The solution

Equation 4 is a differential equation that can be solved by separating variables.

$$\frac{dN}{N \ln(N)} = \beta \delta^\alpha d\delta. \tag{5}$$

Integrating both sides yields:

$$\ln(\ln N) = \frac{\beta}{\alpha + 1} \delta^{\alpha+1} + C, \tag{6}$$

where $C$ is the constant of integration.

A double exponential of both sides provides the solution for $N$:

$$N = N_0 e^{e^{\left(\frac{\beta}{\alpha+1} \delta^{\alpha+1}\right)}}, \tag{7}$$

where $N_0 = e^{e^C}$ is the initial number of earthquakes at $\delta = 0$. The fit reported in Equation 3 provides $\alpha = -0.8 \pm 0.00067$ and $\beta = 0.15 \pm 0.0037$.

## 3.2 Physical Interpretation

The solution in Equation 4 suggests a double exponential relationship between the cumulative number of earthquakes and a power of the ground uplift $\delta$. The parameter $\beta$ represents the proportionality between ground uplift and earthquakes' occurrence rate with $\delta$. Physically, it reflects the efficiency of stress/strain transfer and earthquakes triggering in response to the caldera dynamics (i.e., fluid pressurization, magma transfer, magmatic fluids migration, etc.). High values of $\beta$ indicate a more efficient coupling between deformation and seismicity.

The exponent $\alpha$ introduces non-linearity to account for the efficiency of the triggering mechanism in fluid-induced seismic swarms. This approach is supported by Perfettini and Avouac [35], who demonstrate that aftershock rates in tectonic regions can be directly related to the post-seismic slip rate assuming a velocity-strengthening brittle creep rheology. Furthermore, Lippiello et al. [36] and Lippiello [37] show that the aftershock occurrence rate is proportional to the stress rate in fault zones with velocity-strengthening friction, which justifies the introduction of $\delta$ as a key variable driving earthquake interactions in our model. The nonlinearity introduced by $\alpha$ accounts for the reduced efficiency of earthquake triggering in volcanic swarms compared to tectonic sequences. In fact, a negative value of $\alpha$ reflects the damping effect of the increasing ground uplift on the triggering mechanism (Eq.4). However, being $|\alpha| < 1$, the cumulative number of earthquakes increases with $\delta$ (Eq.7). The physical reason for this observation could be a plastic deformation of the rocks. More precisely, part of the stress responsible for the ground uplift does not generate new earthquakes because it is dissipated in plastic deformation of the surrounding medium.

In general, the presence of the term $N$ in Eq. 4 suggests that the mechanism of the occurrence of earthquakes is an epidemic one (i.e., the system is self-amplifying). This is not a new observation, indeed. In tectonic areas aftershocks tend to occur in a cascading similar to epidemic propagation [21–25] resulting in a hyperbolic decrease of the occurrence rate well known as Omori law [38]. The duration of the sequence is in principle infinite (see, among others, Godano and Tramelli [39]) and, operatively, a sequence ends when the occurrence rate becomes of the same order of the background seismic activity. Conversely, on volcanoes, earthquakes occur in short swarms characterized by the absence of a mainshock [40]. As a consequence, we can imagine that earthquakes trigger one another in a less efficient cascading. In fact, in our model, a negative value of $\alpha$ implies that the triggering efficiency decreases with increasing $\delta$. More precisely, whenever an increase of $\delta$ triggers an earthquake swarm, it itself produces a damping of the occurrence rate testified by the negative value of $\alpha$. This appears to be in very good agreement with the results of Godano et al. [41] who found that the earthquake occurrence rate of volcanic swarms can be described by a decreasing power law exponentially tapered at large times.

The term $\ln(N)$ represents a saturation effect: a sort of competition between earthquakes which makes slower the seismic rate increases. A logarithmic saturation effect has been observed in quasi-fragile disordered media when the distribution of the fracture resistance is wide (see as an example Nukala and Simunovic [42]). This could be the case for highly fractured volcanic areas, more rigid rocks (cap rock) and water saturated ones.

Our model aligns, also, with the understanding that seismic swarms are influenced by complex processes such as fluid diffusion and aseismic deformation [43, 44], which may propagate in a different way compared with stress transfer in tectonic settings [45].

## 4  Conclusions

In this study, we analyzed the relationship between ground uplift and earthquake occurrence during the ongoing unrest at the Campi Flegrei caldera. Our findings reveal that, from 2008 onward, the ground uplift $\delta$ follows two distinct exponential regimes, providing a more robust fit compared to a single power-law model. This highlights the evolving dynamics of the caldera, likely driven by a combination of fluid pressurization and structural adjustments in the upper crust. The most significant result of our analysis is the nonlinear relationship between the cumulative number of earthquakes $N(\delta)$ and the ground uplift $\delta$. Our model, which incorporates the reduced triggering efficiency of volcanic seismic swarms, offers a novel explanation for the observed behavior. The introduction of the parameter $\alpha$ captures the damping effect of increasing $\delta$ on the seismicity rate, consistent with the localized and short-lived nature of volcanic swarms. In terms of the epidemic model for earthquakes' occurrence, a negative value of $\alpha$ implies that the increase of $\delta$ reduces the ability of each parent to generate offspring. Such a result could be the indication of the rocks inelasticity in the area.

Moreover, the logarithmic term in $N$ reveals a saturation effect of the fractures interpretable as a disordered medium with a wide distribution of the resistance to fracture.

These findings challenge traditional views of volcanic unrest, particularly the hypothesis that the upper crust transitions through distinct mechanical regimes (elastic, quasi-elastic, inelastic) during periods of deformation Kilburn [14], Kilburn, De Natale, and Carlino [15], and Kilburn et al. [16]. Instead, our model suggests a continuous process where $\delta$ drives seismicity in a cascading, epidemic-like, modulated by the inherent inefficiency of swarm triggering and by the saturation effect. This result aligns with observations at other volcanoes, where seismic swarms exhibit a power-law occurrence rate tapered by exponential decay [41].

Our work also underscores the potential for using $\delta$ as a predictive variable in volcanic settings. Unlike time-based models, ground uplift directly reflects the physical processes driving seismicity, making it a more reliable indicator of evolving unrest. Future studies should investigate whether this relationship holds across different volcanic systems, especially in calderas, potentially providing a universal framework for understanding volcanic earthquake swarms.

## 5  Data

We use the seismic catalogue recorded by the Istituto Nazionale di Geofisica e Vulcanologia (INGV)-Osservatorio Vesuviano [46] containing 11166 earthquakes with magnitude in the range [-1.1, 4.4] from January 2005 to June 2024 and the soil upwelling recorded at station RITE managed by INGV-Osservatorio Vesuviano [47] in the same period. In the analyses presented in this study we consider only earthquakes with magnitude larger than $m_c$=0.5.

# 6 Methods

The only method we use in the present paper is a non linear fit.

# Declarations

**Funding** This paper was partially supported by the project Non linear models for magma transport and volcanoes generation - project code: P20222B5P9, PRIN 2022 PNRR

**Conflicts of interest/Competing interests.** The authors declare that they have no conflict of interest. This research would be used only for scientific research and would not passed on to third parties.

**Availability of data and material.** The deformation series can be downloaded at https://zenodo.org/record/6389920#.ZD0GCnbMJD8. Whereas the seismic catalogue can be downloaded at the INGV-Osservatorio Vesuviano web site https://terremoti.ov.ingv.it/gossip/flegrei [46].

**Ethical statement** This material is the authors' own original work, which has not been previously published elsewhere.

The paper is not currently being considered for publication elsewhere.

The paper reflects the authors' own research and analysis in a truthful and complete manner.

The paper properly credits the meaningful contributions of co-authors and co-researchers.

The results are appropriately placed in the context of prior and existing research.

All sources used are properly disclosed (correct citation). Literally copying of text must be indicated as such by using quotation marks and giving proper reference.

**Author Contribution** All the authors equally contributed to the manuscript